\documentclass{emulateapj}
\newcommand{\HI}{H${\rm\scriptstyle I}$}
\newcommand{\Ho}{H$^{\rm\scriptstyle o}$}
\newcommand{\HII}{H$^+$}
\newcommand{\Ha}{H$\alpha$}
\newcommand{\Hb}{H$\beta$}
\newcommand{\etal}{\ et al.\ }
\newcommand{\Msun}{M$_\odot$} 
\newcommand{\kms}{$\,\rm {km\,s^{-1}} $} 

\shorttitle{The Source of Ionization along the Magellanic Stream}
\shortauthors{Bland-Hawthorn\etal}
\slugcomment{ApJ Letters, vol. 670 (Dec 1, 2007)}

\begin{document}

\title{The Source of Ionization along the Magellanic Stream}

\author{Joss Bland-Hawthorn$^{1,2}$, Ralph Sutherland$^3$, Oscar Agertz$^4$, Ben Moore$^4$}

\affil{$^1$ Institute of Astronomy, School of Physics, University of Sydney, NSW 2006, Australia} 
\affil{$^2$ Anglo-Australian Observatory, PO Box 296, Epping, NSW 2121, Australia}
\affil{$^3$ Mount Stromlo Observatory, Woden, ACT 2611, Australia}
\affil{$^4$ University of Zurich, Winterthurerstrasse 190, 8057 Zurich, Switzerland}

\begin{abstract}
Since its discovery in 1996, the source of the bright \Ha\ emission (up to 750
mR\footnote{1~Rayleigh (R) =  $10^6/4\pi$ photons cm$^{-2}$ s$^{-1}$ sr$^{-1}$,
equivalent to $5.7\times 10^{-18}$ erg cm$^{-2}$ s$^{-1}$ arcsec$^{-2}$ at \Ha.})
along the Magellanic Stream has remained a mystery. There is no evidence 
of ionising stars within the \HI\ stream, and the extended hot halo is far too 
tenuous to drive strong shocks into the clouds. We now present a hydrodynamical 
model that explains the known properties of the \Ha\  emission and provides 
new insights on the lifetime of the Stream clouds. The upstream clouds are 
gradually disrupted due to their interaction with the hot halo gas. 
The clouds that follow plough into gas ablated from the upstream clouds, 
leading to shock ionisation at the leading edges of the downstream clouds. Since 
the following clouds also experience ablation, and weaker \Ha\ (100$-$200 mR)
is quite extensive, a disruptive cascade must be operating along much of the Stream. 
In our model, the clouds are evolving on timescales of 100$-$200~Myr, such that the
Stream must be replenished by the Magellanic Clouds at a fairly constant rate.
The ablated material falls onto the Galaxy as a warm drizzle which suggests that diffuse ionized gas 
at 10$^4$K may be an important constituent of galactic accretion. The observed \Ha\ emission 
provides a new constraint on the rate of disruption of the Stream and, consequently, the infall 
rate of metal-poor gas onto the Galaxy. When the ionized component of the Stream is fully 
accounted for, the rate of gas accretion is 0.4 M$_\odot$ yr$^{-1}$, roughly twice the rate
deduced from \HI\ observations alone.
\end{abstract}


\section{Introduction}
How do galaxies get their gas? This important question has never been fully
answered, either through observation or through numerical simulation. \HI\
observations of the nearby universe suggest that galaxy mergers and collisions
are an important aspect of this process (Hibbard \& van Gorkom 1996), but tidal interactions do not 
guarantee that the gas settles to one or other galaxy. The most spectacular interaction
phenomenon is the Magellanic \HI\ Stream that trails from the LMC-SMC system 
(10:1 mass ratio) in orbit about the Galaxy. Since its discovery (Wannier \& 
Wrixon 1972; Matthewson et al 1979), there have been repeated attempts to explain the Stream 
in terms of tidal and/or viscous forces (q.v. Mastropietro et al. 2005; Connors, Kawata \& 
Gibson 2005). Indeed, the Stream has become a benchmark against which to judge 
the credibility of N-body$+$gas codes in explaining gas processes in galaxies. A fully 
consistent model of the Stream continues to elude even the most sophisticated codes.

Here, we demonstrate that \Ha\ detections along the Stream (Weiner \& Williams 1996; Putman 
et al 2003) are providing new insights on the present state and evolution of the \HI\ gas. At a 
distance of $D\approx 55$~kpc, the expected \Ha\ signal excited by the cosmic and Galactic 
UV backgrounds are about 3~mR and 25~mR respectively (Bland-Hawthorn \& Maloney 1999, 
2002), significantly lower than the mean signal of 100$-$200 mR, and much 
lower than the few bright detections in the range $400-750$ mR (Weiner, Vogel \& Williams 
2002). This signal cannot have a stellar origin since repeated attempts to detect stars along the 
Stream have failed (e.g. Ostheimer, Majewski \& Kunkel 1997).

Some of the Stream clouds exhibit compression fronts and head-tail morphologies 
(Br\"uns et al 2005) and this is suggestive of confinement by a tenuous external medium. 
But the cloud:halo density ratio ($\eta = \rho_c/\rho_h$) necessary for confinement can be orders 
of magnitude {\it larger} than that required to achieve shock-induced \Ha\ emission 
(e.g. Quilis \& Moore 2001). Indeed,  the best estimates of the halo density at the distance of the 
Stream ($\rho_h \sim 10^{-4}$ cm$^{-3}$;  Bregman 2007) are far too tenuous to induce strong 
\Ha\ emission at a cloud face. It is therefore surprising to discover that the brightest \Ha\ 
detections lie at the leading edges of \HI\ clouds (Weiner\etal\  2002) and thus appear to 
indicate that shock processes are somehow involved.
 
We now present a model that goes a long way towards explaining the \Ha\ mystery.
The basic premise is that a tenuous external medium not only confines clouds, but
also disrupts them with the passage of time. The growth time for Kelvin-Helmholtz (KH) instabilities 
is given by $\tau_{\rm KH} \approx \lambda \eta^{0.5}/v_h$ where $\lambda$ is the wavelength of the 
growing mode, and $v_h$ is the apparent speed of the halo medium ($v_h\approx 350$ km s$^{-1}$;
see \S 2). 
For cloud sizes of 
order a few kiloparsecs and $\xi \approx 10^4$, the KH timescale can be much less than an 
orbital time ($\tau_{\rm MS} \approx 2\pi D/v_h \approx 1$ Gyr).
Once an upstream cloud becomes disrupted, the fragments are slowed with respect to the 
LMC-SMC orbital speed and are subsequently ploughed into by the following
clouds. In \S 2, the new hydrodynamical models are described and the results are
presented in \S 3.  In \S 4, we discuss the implications of our model and suggest avenues for 
future research.

\section{A new hydrodynamical model}

There have been many attempts to understand how gas clouds interact with an ambient medium 
(Murray, White \& Blondin 1993; Klein, McKee \& Colella 1994). In order to capture the evolution of a 
system involving instabilities with large density gradients correctly, grid based methods 
(Liska \& Wendroff 1999; Agertz\etal\ 2007) are favoured over other schemes (e.g. Smooth Particle Hydrodynamics). 
We have therefore investigated the dynamics of the Magellanic Stream with two 
independent hydrodynamics codes, {\it Fyris} (Sutherland 2007) and {\it Ramses} (Teyssier 2002), that 
solve the equations of gas dynamics with adaptive mesh refinement. The results shown here are 
from the {\it Fyris} code because it includes non-equilibrium ionization, but we get comparable 
gas evolution from either code\footnote{Further details on the codes and comparative simulations are
provided at http://www.aao.gov.au/astro/MS.}.

The brightest emission is found along the leading edges of clouds MS~II, III and 
IV with values as high as 750 mR  for MS II. The \Ha\ line 
emission is clearly resolved at $20-30$ km s$^{-1}$ FWHM, and shares the same radial velocity 
as the \HI\ emission within the measurement errors (Weiner et al 2002; G. Madsen 2007, personal
communication). This provides an important constraint on the
physical processes involved in exciting the Balmer emission.

In order to explain the \Ha\ detections along the Stream, we concentrate our efforts on the 
disruption of the clouds labelled MS I$-$IV (Br\"uns\etal\ 2005). The Stream is trailing the 
LMC-SMC system in a counter-clockwise, near-polar orbit as viewed from the Sun. 
The gas appears to extend from the LMC dislodged through tidal disruption
although some contribution from drag must also be operating (Moore \& Davis 1994). Recently, 
the Hubble Space Telescope has determined an orbital velocity of 378$\pm$18 km s$^{-1}$ 
for the LMC. While this is higher than earlier claims, the result has been confirmed by 
independent researchers (T. Pryor, personal communication). Besla et al (2007) conclude 
that the origin of the Stream may no longer be adequately explained with existing numerical 
models. The Stream velocity along its orbit must be comparable to the 
motion of the LMC; we adopt a value of $v_{\rm MS} \approx 350$ km s$^{-1}$.


\begin{figure}[htbp]
\begin{center}
\caption{The initial fractal distribution of \HI\ at 20~Myr, shown in contours, before the wind action has taken hold. The upper figure is in the $x-z$ frame as seen from above; the lower figure is the projected distribution on the plane 
of the sky ($x-y$ plane). Both distributions are integrated along the third axis. The logged \HI\ contours correspond to 18.5 (dotted), 19.0, 19.5, 20.0, and 20.5 (heavy) cm$^{-2}$. The greyscale shows weak levels of \Ha\ along
the Stream where black corresponds to 300~mR.}
\label{default}
\end{center}
\end{figure}

Here we employ a 3D Cartesian grid with dimensions $18\times 9 \times9$~kpc [$(x,y,z) = 
(432, 216, 216)$ cells] to model a section of the Stream where $x$ is directed along the Stream arc and
the $z$ axis points towards the observer. The grid is initially filled with two gas components. 
The first is a hot thin medium representing the halo corona.  If we adopt a rigorously isothermal halo for 
the Galaxy, with parameters from Battaglia\etal\ (2005), the virial temperature at a radial distance of 55 kpc is 
$T_h =1.75\times 10^6$ K. Embedded in the hot halo is (initially) cold \HI\ material with a total \HI\ mass of 
$3\times 10^7$ M$_\odot$. The cold gas has a fractal distribution and is initially confined to a cylinder with 
a diameter of 4 kpc and length 18 kpc (Fig. 1); the mean volume and column densities are 
0.02 cm$^{-3}$ and $2\times 10^{19}$ cm$^{-2}$ respectively. The 3D spatial power spectrum 
($P(k)\propto k^{-5/3}$) describes a Kolmogorov turbulent medium with a minimum wavenumber $k$ 
corresponding to a spatial scale of 2.25 kpc, comparable to the size of observed clouds along the Stream.
 
A key parameter of the models is the ratio of the cloud to halo pressure, $\xi = P_c/P_h$. 
If the cloud is to survive the impact of the hot halo, then $\xi \ga 1$. A shocked cloud is 
destroyed in about the time it takes for the internal cloud shock to cross the cloud, during which time 
the cool material mixes and ablates into the gas streaming past. Only massive clouds with dense cores
can survive the powerful shocks. An approximate lifetime for a spherical cloud of diameter $d_c$ is
\begin{equation}
\tau_{\rm c} = 60  (d_c/2 {\rm\; kpc})^{-1} ( v_{\rm h}/350 {\rm\; km s}^{-1})^{-1} (\eta/100)^{0.5} \,  {\rm Myr}.
\end{equation}
For $\eta$ in the range of 100$-$1000, this corresponds to 60$-$180 Myr for individual clouds. 
With a view to explaining the \Ha\ observations, we focus our simulations on the lower end of this range.

For low $\eta$, the density of the hot medium is $n_h=2\times 10^{-4}$ cm$^{-3}$.
The simulations are undertaken in the frame of the cold \HI\ clouds, so the halo gas is given an initial 
transverse velocity of 350\kms. The observations reveal that the mean \Ha\ emission has a slow trend
along the Stream which requires the Stream to move through the halo at a small angle of
attack (20$^o$) in the plane of the sky (see Fig. 1). Thus, the velocity of the hot gas as seen by the Stream is
$(v_x,v_y)$ $=$ $(-330,-141)$\kms. The adiabatic sound speed of the halo gas is 
200\kms, such that the drift velocity is mildly supersonic (transsonic), with a Mach number of 1.75.

A unique feature of the {\it Fyris} simulations is that they include non-equilibrium cooling through 
time-dependent ionisation calculations (cf. Rosen \& Smith 2004). When shocks occur within the 
inviscid fluid, the jump shock conditions are solved across the discontinuity. This allows us to 
calculate the Balmer
emission produced in shocks and additionally from turbulent mixing along the Stream (e.g. Slavin\etal\ 1993). 
We adopt a 
conservative value for the gas metallicity of [Fe/H]=-1.0 (cf. Gibson\etal\ 2000); a higher value 
accentuates the cooling and results in denser gas, and therefore stronger \Ha\ emission along the
Stream.


\begin{figure}[htbp]
\begin{center}
\caption{The dependence of the evolving fractions of \Ho\ and \HII\ on column density as the shock cascade
progresses. The timesteps are 70 (red), 120 (magenta), 170 (blue), 220 (green) and 270 Myr (black).
The lowest column \HI\ becomes progressively more compressed with time but the highest column \HI\ is 
shredded in the cascade process; the fraction of ionized gas increases with time. The pile-up of electrons
at low column densities arises from the x-ray halo.}
\label{default}
\end{center}
\end{figure}

\section{Results}

The results of the simulations are shown in Figs. $2-4$; we provide animations of
the disrupting stream at http://www.aao.gov.au/astro/MS. In our
model, the fractal Stream experiences a ``hot wind" moving in the
opposite direction. The sides of the Stream clouds are
subject to gas ablation via KH instabilities due to the reduced
pressure (Bernouilli's theorem). The ablated gas is slowed
dramatically by the hot wind and is transported behind the cloud.
As higher order modes grow, the fundamental mode associated
with the cloud size will eventually fragment it. The ablated gas
now plays the role of a ``cool wind" that is swept up by the pursuing
clouds leading to shock ionization and ablation of the downstream
clouds. The newly ablated material continues the trend along the
length of the Stream. The pursuing gas cloud transfers momentum
to the ablated upstream gas and accelerates it;
this results in Rayleigh-Taylor (RT) instabilities, especially at the stagnation point in the 
front of the cloud. We rapidly approach a nonlinear regime where the KH and RT
instabilities become strongly entangled, and the internal motions become highly turbulent.
The simulations track the progression of the shock fronts as they propagate into the cloudlets.

In Fig. 2, we show the predicted conversion of neutral to ionized hydrogen due largely to cascading
shocks along the Stream. The drift of the peak to higher columns is due to the shocks eroding away the outer
layers, thereby progressing into increasingly dense cloud cores. The ablated gas drives a shock 
into the \HI\ material with a shock speed of $v_s$ measured in the cloud 
frame. At the shock interface, once ram-pressure equilibrium is reached, we find
$v_s \approx v_{\rm h} \eta^{-0.5}$. In order to produce significant \Ha\ emission, $v_s \ga 35$\kms\ such
that $\eta \la 100$. In Fig.~3, we see the steady rise in \Ha\ emission along the Stream, reaching $100-200$
mR after 120~Myr, and the most extreme observed values after 170~Myr. The power-law
decline to bright emission measures is a direct consequence of the shock cascade. The shock-induced 
ionization rate is $1.5\times 10^{47}$ phot s$^{-1}$ kpc$^{-1}$. The predicted luminosity-weighted line 
widths of 20\kms\ FWHM (Fig. 3, inset) are consistent with the \Ha\ kinematics.
In Fig. 4, the \Ha\ emission is superimposed onto the projected \HI\ emission: much of it lies at 
the leading edges of clouds, although there
are occasional cloudlets where ionized gas dominates over the neutral column. Some of the brightest 
emission peaks appear to be due to limb brightening, while others arise from chance alignments.

The simulations track the degree of turbulent 
mixing between the hot and cool media brought on by KH instabilities (e.g. Kahn 1980). The turbulent 
layer grows as the flow develops, mixing up hot and cool gas at a characteristic temperature of 
about 10$^4$K. In certain situations, a sizeable \Ha\ luminosity can be generated (e.g. Canto \& 
Raga 1991) and the expected line widths are comparable to those observed in the Stream (\S 2). 
Indeed, the simulations reveal that the fractal clouds develop a warm ionized skin along the entire
length of the Stream. But the characteristic \Ha\ emission (denoted by the shifting peak in Fig. 3)
is comparable to the fluorescence excited by the 
Galactic UV field (Bland-Hawthorn \& Maloney 2002). We note with interest that narrow Balmer lines can
arise from pre-cursor shocks (e.g. Heng \& McCray 2007), but these require conditions that are unlikely 
to be operating along the Stream.


\begin{figure}[htbp]
\begin{center}
\caption{The evolving distribution of projected \Ha\ emission as the shock cascade progresses. 
The timesteps are explained in Fig. 2.
The extreme emission measures increase with time and reach the observed mean values after 120~Myr.
The mean and peak emission measures along the Stream are indicated, along with the approximate 
contributions from the cosmic and Galactic UV backgrounds.
{\bf Inset:} The evolving \Ha\ line width as the shock cascade progresses; the velocity scale is with respect
to the reference frame of the initial \HI\ gas. The solid lines are flux-weighted line profiles; the dashed lines are 
volume-weighted profiles that reveal more extreme kinematics at the lowest densities.}
\label{default}
\end{center}
\end{figure}

\begin{figure}[htbp]
\begin{center}
\caption{The predicted \HI\ (contours) and \Ha\  (greyscale) distributions after 120~Myr; the intensity levels 
are given in Fig. 1. The angle of attack in the [x,y,z] coordinate frame is indicated. The \Ha\ emission is 
largely, but not exclusively, associated with dense \HI\ gas.} 
\label{default}
\end{center}
\end{figure}

\section{Discussion}

We have seen that the brightest \Ha\ emission along the Stream can be understood in terms of shock 
ionization and heating in a transsonic (low Mach number) flow. For the first time, the Balmer emission (and 
associated emission lines) provides diagnostic information at any position along the Stream that is 
independent of the \HI\  observations. Slow Balmer-dominated shocks of this kind (e.g. Chevalier
\& Raymond 1978) produce partially ionized media where a significant fraction of the \Ha\ emission
is due to collisional excitation. This can lead to Balmer decrements (\Ha/\Hb\ ratio) in excess of 4, i.e. 
significantly enhanced over the pure recombination ratio of about 3,
that will be fairly straightforward to verify in the brightest regions of the Stream.

The shock models predict a range of low-ionization emission
lines (e.g. OI, SII), some of which will be detectable even though suppressed by the low gas-phase
metallicity. There are likely to be EUV absorption-line diagnostics through the shock interfaces
revealing more extreme kinematics (Fig. 3, inset), but these detections (e.g. OVI) are only possible 
towards fortuitous background sources (Sembach\etal\ 2001; Bregman 2007). The predicted 
EUV/x-ray emissivity from the post-shock regions is much too low to be detected in emission.

The characteristic timescale for large changes is roughly 100$-$200 Myr,
and so the Stream needs to be replenished by the outer disk of the LMC at a fairly constant rate
(e.g. Mastropietro\etal\ 2005). The timescale can be extended with larger $\eta$ values (see equation (1)), 
but at the expense of substantially diminished \Ha\ surface brightness. In this respect, we consider $\eta$ 
to be fairly well bounded by observation and theory.

What happens to the gas shedded from the dense clouds? Much of the diffuse gas will
become mixed with the hot halo gas suggesting a warm accretion towards the inner Galactic halo. 
If most of the Stream gas enters the Galaxy via this process, the derived gas accretion rate is
$\sim 0.4$\Msun\ yr$^{-1}$. The higher value compared to \HI\ (e.g. Peek\etal\ 2007) is due to the 
gas already shredded, not seen by radio telescopes now. In our model, the HVCs
observed today are unlikely to have been dislodged from the Stream by the process
described here. These may have come from an earlier stage of the LMC-SMC interaction with the 
outer disk of the Galaxy.

The ``shock cascade'' interpretation for the Stream clears up a nagging uncertainty about the 
\Ha\ distance scale for high-velocity clouds.  Bland-Hawthorn et al (1998) first showed that
distance limits to HVCs can be determined from their observed \Ha\ strength due to 
ionization by the Galactic radiation field, now confirmed by clouds with reliable distance 
brackets from the stellar absorption line technique (Wakker 2001).  HVCs
have smaller kinetic energies compared to the Stream clouds, and
their interactions with the halo gas are not expected to produce significant shock-induced 
or mixing layer \Ha\ emission, thereby supporting the use of \Ha\ as a crude distance indicator.

Here, we have not attempted to reproduce the \HI\ observations of the Stream in detail. 
This is left to a subsequent paper where we explore a larger parameter space and include a more 
detailed comparison with the \HI\ and \Ha\ power spectrum, inter alia. We introduce additional physics, 
in particular, the rotation of the hot halo, a range of Stream orbits through the halo gas, the 
gravitational field of the Galaxy, and 
so on. 

If we are to arrive at a satisfactory understanding of the Stream interaction with the halo, future 
deep \Ha\ surveys will be essential. It is plausible that current \Ha\ observations are still missing
a substantial amount of gas, in contrast to the deepest \HI\ observations. We can compare
the particle column density inferred from \HI\ and \Ha\ imaging surveys. The limiting
\HI\ column density is about  $N_H \approx \langle n_H \rangle L \approx 10^{18}$ cm$^{-2}$
where $\langle n_H \rangle$ is the mean atomic hydrogen density, 
and $L$ is the depth through the slab. By comparison, the \Ha\ 
surface brightness can be expressed as an equivalent emission measure, 
$E_m \approx \langle n_e^2 \rangle L \approx \langle n_e\rangle N_e$. Here
$n_e$ and $N_e$ are the local and column electron density. 
The limiting value of $E_m$ in \Ha\ imaging is about 100 mR,
and therefore $N_e \approx 10^{18}/\langle n_e \rangle$ cm$^{-2}$.
Whether the ionized and neutral gas are mixed or distinct, we can
hide a lot more ionized gas below the imaging threshold for a fixed $L$, particularly
if the gas is at low density ($\langle n_e \rangle \ll 0.1$ cm$^{-3}$).
A small or variable volume filling factor can complicate this picture but, 
in general, the ionized gas still wins out because of ionization of low 
density \HI\ by the cosmic UV background (Maloney 1993).
In summary, even within the constraints of the cosmic microwave background
(see Maloney \& Bland-Hawthorn 1999),
a substantial fraction of the gas can be missed if it occupies
a large volume in the form of a low density plasma.

\acknowledgments  JBH is indebted to the University of Zurich for organizing the Saas Fee lectures 
this year at M\"urren in the Swiss Alps, a setting that provided the inspiration for the new work. 
We wish to thank the referee for insightful and helpful comments.
We are indebted to G. Besla, G. Madsen and T. Pryor for important data in advance of publication.
JBH is supported by a Federation Fellowship through the Australian Research Council (ARC).
RSS and JBH gratefully acknowledge ARC grant DP0664434 that supports certain aspects of this work.

\end{document}